# Super-resolution by converting evanescent waves in microsphere to propagating and transfer function from its surface to nano-jet

Y. Ben-Aryeh

Physics Department, Technion- Israel-Institute of Technology, Haifa, 32000, Israel

e-mail: phr65yb@physics.technion.ac.il

**Abstrac**t

The EM waves transmitted through a thin object with fine structures is observed, by microsphere located above the object. While the EM waves include both "evanescent" and "propagating" waves, the high resolution is obtained by the evanescent ones, including the information on the fine structures of the object. Description of this process is divided into two parts:  a) The super-resolution is analyzed by using Helmholtz equation for the evanescent waves transmitted from the object to the microsphere surface.  b) Using boundary condition, the electric fields on the inner surface of the microsphere includes both the evanescent and propagating waves.  The transmission of these waves to a nano-jet is produced by a transfer function, including convolution between the spatial modes of the evanescent waves with those of the microsphere, which increases the conversion of the evanescent waves to propagating waves, and thus increase the resolution.

## 1. Introduction

Any microscopic image can be magnified by using a microscope. But observing sub-wavelength structures is difficult because of the *Abbe diffraction limit* [1] by which light with wavelength $\lambda$ travelling in medium with refractive index $n$ and aperture angle $\theta$ will make a spot with radius

$$d = \frac{2(n\sin\theta)}{\lambda} \quad . \tag{1}$$

The term $n\sin\theta$ is called the numerical aperture $(NA)$ and the Abbe limit is of order $\lambda/2$.

I analyze in the present article the mechanism by which evanescent waves incident on a microsphere are converted into propagating waves, and by such conversion high resolution is obtained in the image by the microsphere, which is much beyond the Abbe limit.

## 2. Methods

The use of evanescent waves to increase the resolution beyond the Abbe limit can be related to Helmholtz equation [2]. In homogenous medium this equation is given by

$$(nk_0)^2 = k_x^2 + k_y^2 + k_z^2 \tag{2}$$

where $k_0 = 2\pi/\lambda_0$, $\lambda_0$ is the wavelength in vacuum, $k_x, k_y, k_z$ are the wavevector components. The evanescent waves satisfying the relation:

$$k_x^2 + k_y^2 > (nk_0)^2 \quad . \tag{3}$$

are arriving at the microsphere with imaginary $k_z$. The increase of the component of the wavevector $\vec{k}$ in the plane $x, y$ decreases the "effective" wavelength in this plane, and thus increases the resolution.

As described in Figure 1, a dielectric microsphere with radius R, and a refractive index $n_2$ is located above a thin object at contact point O. The medium between the object and the microsphere has a refractive index $n_1$. Parallel EM waves are transmitted in direction perpendicular to thin planar object, which may be transmitted as "propagation" and "evanescent". But the increase of resolution by the microsphere is related to conversion of evanescent waves to propagating waves.

We analyze the conversion of evanescent waves to propagating waves at a point P which is on the microsphere surface. The incident and transmitted angle for the EM waves transmitted into the microsphere at point P are given as $\Theta_I$ and $\Theta_T$, respectively. At this point, EM waves with wavevectors $k_x, k_y$ are arriving at the microsphere where $k_z$ is imaginary. The increase of the component of the wavevector $\vec{k}$ in the plane $x, y$ decreases the "effective" wavelength in this plane, and thus increases the resolution. But the evanescent waves decay in the perpendicular z direction, so that to "capture" the fine structure which is available in the evanescent waves we need that the point P will be near the contact point O, so that its perpendicular distance to the object will be of a wavelength order. According to geometric optics the microsphere has a spherical symmetry under rotation around the z axis, which connects the center of the microsphere at point C, with the contact point O. The line connecting the center point C with point P has an angle $\Theta_I$ with the symmetric z axis.

As shown in Figure 1, the EM waves transmitted through the microsphere are converging into photonic jet (PJ), where its role in producing the high resolution is controversial. The original work in obtaining high resolution in microsphere imaging was made by Zengo Wang et al. [3]. That analysis was made with virtual image as follows from the geometric optics description. Since this time very large number of papers were published on various effects in the microsphere system, In Figure 1, we describe a real, image which is produced by using high-refractive index microsphere (e. g. [4, 5]).

The focusing of light in the microsphere system is concentrated in the nano-jet sub-diffraction region, that does not obey the classical laws of geometrical optics. but might be explained by diffraction effects, known as photonic jet (PJ) (e.g., [6,7]). Exact solutions for non-diffraction beams might be related to the central part of the photonic nano-jet [8,9]. Rigorous Mie theory predicts the interaction of light with spherical particles and this theory was used to describe various properties of the PJ's, produced by the microsphere system [10]. The exact use of the Mie theory is usually done by numerical calculations, as it is obtained by sum of many terms that do not give analytical result. Using Mie theory optical resonances in microsphere photonic nano-jets were observed [11]. We analyzed the properties of evanescent waves, produced by plane EM waves transmitted through nano-corrugated-metallic thin film, which includes the information on its fine-structures [12]. A microsphere located above the metallic surface collects the evanescent waves, which are converted to propagating waves. The magnification of the nano-structure images is explained by optical description, but the high resolution is related to the evanescent wave description. Such approach for explaining the high resolutions obtained by microspheres was developed by using complex Snell's law [13]. Very high resolutions by microspheres were reported also in other works (e. g. [14,15]).

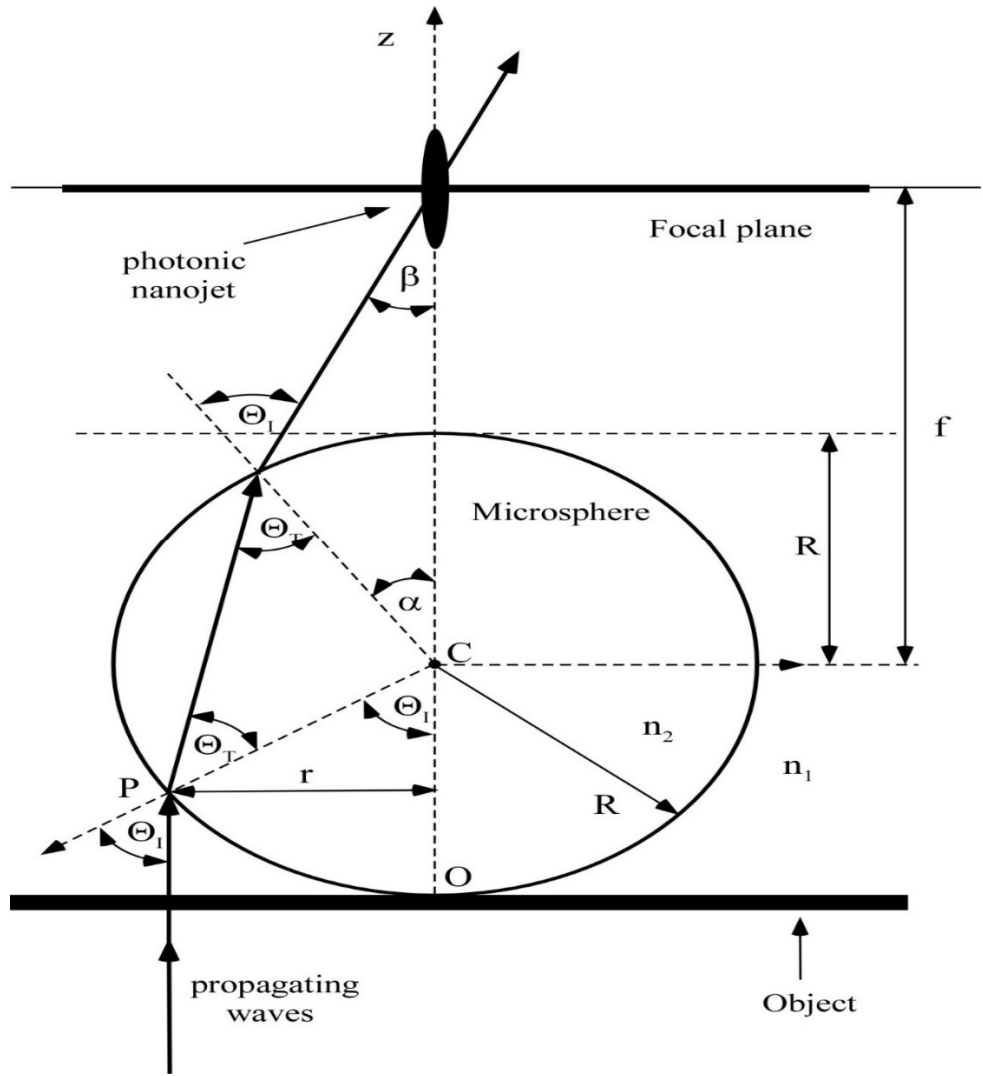

 **Caption for Figure 1**

Propagating EM waves are transmitted through a thin object with a certain structure, whose image is produced by evanescent waves, arriving at the microsphere surface at point P, for example. The EM transmitted through the microsphere are converging into a non-diffraction region, known as photonic jet (PJ).

Maslov and Astratov [16,17] studied the origin of super-resolution in microsphere assisted imaging. They claimed that the imaging of the nano-jet cannot give good explanation to the high resolutions obtained in the microsphere system. They claimed also that the evanescent waves source cannot explain the high resolutions obtained by the microsphere system since its effect is too weak. They suggested to use a direct approach to Maxwell equations including optical principles such as point-spread functions and reciprocity in micro spherical imaging. I claim, however, that there are strong arguments that evanescent waves should be included in the theories about high resolutions: a) The propagating distance between the object and the microscope must be small enough to reveal the nanometric features. To generate super-resolution over large areas of the sample we need to attach the microsphere to a frame, which is scanned on the sample in step-by step fashion [18]. b) There is a dependence of the super-resolution on the radius of the microsphere and its index of refraction which is in good correspondence with the evanescent wave properties [13]. c) The super-resolution is related to information theory where the information on the fine structures of the object area is available in the evanescent waves. Magnification of the image and optical transforms cannot introduce high resolution, if the information on the object fine structures is not available in such transforms. I find, however. that there is an important mechanism which enhances the transformation of evanescent waves to propagating waves after the microsphere surface, which is beyond classical geometric optics (beyond the use of complex Snell's law [13]). To explain this phenomenon, it will be helpful to compare the high resolution obtained by the microsphere with the high resolution obtained by other systems, which is done as follows.

Experiments have shown that a thin film metallic grating with arrays of subwavelength holes can transmit light at certain frequencies, which is order of magnitude larger than the light intensity incident on the area of these holes. Most investigators agree that these experiments conducted originally by Ebessen et al. [19,20], are related to coupling the light with surface plasmons. I explained these phenomena as transmission enhancement by conversion evanescent waves, entering the small holes to propagating waves, due to *convolution of the evanescent spatial modes with the plasmons spatial modes,* producing high spatial wavevectors, with very small "effective" wavelength [21]. Such effects were described in a similar way by relating the 'tunnelling' of evanescent waves to propagating waves due to the convolution of the high spatial frequencies of the source with those of the detector [22].

The super-resolution obtained by the microsphere is like the field of scanning near-field optical microscopy (SNOM), where resonant effect with tip detector enhances the super-resolution. One should consider that in the dielectric sphere we don't have plasmons, but we have other coupling mechanisms. We follow in the present analysis the idea that the evanescent wavevectors are coupled to the microsphere, by resonances produced by Mie theory analysis [11]. Such Mie theory is used also for a good description of the nano-jet [10]. The enhancement of conversion of evanescent waves to propagating waves, *after the transmission through the microsphere surface*, is obtained by convolution between the evanescent spatial modes and the transfer function of the microsphere modes [2]. The idea of using such convolution was suggested already in previous article [23]. My approach to microsphere high resolution will be developed in the next Section, by using two steps: 1) In the first step we use the Helmholtz equation by which large wave vectors are produced above the object (see Eq. (3)), producing high resolutions. 2) In the second step there is, after the microsphere surface, an enhancement for the conversion of evanescent waves to propagating waves due to coupling between evanescent waves and resonances, produced in the microsphere, for example by Mie theory. In the second stage we do not use Helmholtz equation, and the above coupling is described, as convolution between the spatial modes of the evanescent waves and those of the microsphere modes, described by the *transfer function*.

## 3. An analysis for the super resolution, obtained in the microsphere system

We develop the analysis for the microsphere system into two parts: a) In the first part, we describe the propagation of evanescent waves, produced on thin planar object with fine structures, to the microsphere surface. In this stage high resolutions of the image are obtained related to the use of Helmholtz equation.b) In the second part, we consider the propagation, of both evanescent and propagating waves, produced in the inner surface of the microsphere to the nano-jet.This propagation is described by the convolution of these EM fields with the microsphere modes described by a transfer function [2] which is related to Mie theory, but it will be more convenient to use it as experimental function.

**The use of Helmholtz equation for getting high resolutions by evanescent waves**

Let us assume that the planar surface of an object is given by z=0, and the EM field in this plane is given by the Fourier transform

$$U(x,y) = \int_{-\infty}^{\infty}\int_{-\infty}^{\infty} u(k_x, k_y)\exp\left[-i(k_x x + k_y y)\right] dk_x dk_y \quad . \tag{4}$$

Then, the EM waves propagating from the planar surface of the object into homogenous medium in the space $z > 0$ with a refractive index $n_1$ is given by:

$$U(x,y,z>0) = \int_{-\infty}^{\infty}\int_{-\infty}^{\infty} u(k_x, k_y)\exp\left[-i(k_x x + k_y y + k_z z)\right] dk_x dk_y \quad . \tag{5}$$

Substituting Eq. (5) into the Helmholtz equation:

$$(\Delta^2 + k^2)U(\vec{r}) = 0 \tag{6}$$

we get:

$$\left[k^2 - (k_x^2 + k_y^2 + k_z^2)\right] U(x,y,z>0) = 0 \quad . \tag{7}$$

Under the condition $k^2 < k_x^2 + k_y^2$, $k_z$ is imaginary, and for such case we get the evanescent wave solution:

$$U(x,y,z>0) = U(x,y,z=0)\exp(-\gamma z) ;$$
$$\gamma = \sqrt{k_x^2 + k_y^2 - k^2} \tag{8}$$

For evanescent waves, there is a decay of the wave in the z direction. The resolution obtained by the evanescent waves is limited by the lateral component of the wavelength given by:

$$\lambda_T = \frac{2\pi}{k_T} = \frac{2\pi}{\sqrt{k_x^2 + k_y^2}} \quad . \tag{9}$$

As the evanescent waves satisfy the equation $k^2 = k_x^2 + k_y^2 - \gamma^2$, then Eq. (9) can be written as

$$\lambda_T(evan.) = \frac{2\pi}{\sqrt{k^2 + \gamma^2}} \tag{10}$$

The *minimal* value of $\lambda_T$ for propagating waves is given by:

$$\lambda_{T,\min}(prop.) = \frac{2\pi}{k} \tag{11}$$

since the minimum is obtained when $\vec{k}$ is in the x, y plane.

As by the Abbe limit the resolution is of order $\lambda/2$, the increase of resolution by using evanescent waves is given by:

$$F = \lambda_T(prop.)/\lambda_T(evan.) = \sqrt{\frac{k^2 + \gamma^2}{k^2}} = \sqrt{1 + \frac{\gamma^2}{n_1^2 k_0^2}} \quad . \tag{12}$$

The distance of the point P from the planar object is given by $h = R(1 - \cos\theta_I)$ (see Figure 1). For decay constant $\gamma$ of the evanescent waves, represented in unit $\gamma/n_1 k_0$, the decay of the evanescent wave at point at point P, after transversing the distance $h$, is given by

$$\exp(-\gamma h) = \exp\left[-\frac{\gamma}{n_1 k_0} n_1 k_0 R(1 - \cos\theta_I)\right]$$
$$= \exp\left[-\frac{\gamma}{n_1 k_0} 2\pi (1 - \cos\theta_I)\frac{R}{\lambda}\right] \quad . \tag{13}$$

In our paper on complex Snell's law, Table 1. we demonstrated the decay of the evanescent wave as function of the parameters $\frac{\gamma}{n_1 k_0}$ and $(1 - \cos\theta_I)$ for a certain $\frac{R}{\lambda}$, representing the radius of the microsphere as equal to the product of the wave length with a small number [10]. (Notice that in [13] we used the notation $\alpha$ instead of $\theta_I$). We find that this decay increases very much by increasing $\theta_I$, so that only a small distance around the contact point O is efficient in obtaining the high resolution by evanescent waves.

## Microsphere imaging by a transfer function from the microsphere surface to the nano-jet

Let us assume that the EM field beyond the microsphere surface (of evanescent wave in a small region around the contact point O, plus propagating wave in a larger region) is given by:

$$E'(x, y) = E_{\tan}(x, y) G(x, y) \tag{14}$$

where $E_{\tan}(x,y)$ is the component of EM field, over the microsphere surface, which is tangent to the microsphere surface, and $G(x,y)$ is the 'transfer function' by the microsphere which can be expressed by Fourier transform as [2]:

$$G(k_x,k_y) = \int_{-\infty}^{\infty}\int_{-\infty}^{\infty} G(x,y)\exp\left[-i(k_x x + k_y y)\right]dxdy \qquad (15)$$

The EM field $E_{\tan}(x,y)$, which is tangent to the microsphere surface, at the points x, y on its surface, is preserved in the transmission through the surface [13], and it includes both the evanescent waves and the propagating waves. One should notice that $G(k_x,k_y)$ should include both evanescent waves for which $k_x^2 + k_y^2 > k^2$, and propagating waves for which $k_x^2 + k_y^2 < k^2$

Due to the small distance between the point x, y and the contact point O, the evanescent field $E_{\tan,evan}(x,y)$ is approximately parallel to the x, y plane. This EM field is smaller from the evanescent EM field on the thin object, by the factor given approximately by Eq. (13).

The Fourier amplitude $A(k_x,k_y)$ is described by the Fourier inverse of $E_{\tan}(x,y)$, which is given by:

$$A(k_x,k_y) = \int_{-\infty}^{\infty}\int_{-\infty}^{\infty} E_{\tan}(x,y)\exp\left[-i(k_x x + k_y y)\right]dxdy \qquad (16)$$

where $k_x, k_y$ are the wavevectors in the x and y directions. The convolution of the spatial wavevectors $G(k_x,k_y)$ and $A(k_x,k_y)$ is given by [2]:

$$E'(k_x,k_y) = \int_{-\infty}^{\infty}\int_{-\infty}^{\infty} A(k'_x,k'_y)G(k_x - k'_x, k_y - k'_y)dk'_x dk'_y \qquad (17)$$

The convolution given by Eq. (17) includes spatial wavevectors $k_x - k'_x$ and $k_y - k'_y$ which are very large, so that the corresponding wavelengths are reduced to very low values. For example, if approximately, $k_x - k'_x = kn$ where n is a large integer, then the effective wavelength will be reduced to $\lambda/n$ (see the analysis in [23]). We use the convolution effect described by Eq. (17), to obtain enhancement of the conversion of evanescent waves to propagating waves by reducing the effective wavelength. Although the modulation of the EM waves by evanescent waves is small relative to the total intensity, such modulation is effective in producing the high resolution. The spreading of the spatial modes by convolution is like the point spread function used in [16,17], but one should not ignore the super resolution obtained for the object in the first stage of the imaging process, which is related to the use of Helmholtz equation.

The above reduction of the effective wavelength by convolution, was described also as a quantum effect where entanglement between n photons reduces the effective wavelength to $\lambda/n$ [24,25]. Such entanglement remains also after the transformation to propagating waves.

## 4. Discussion and conclusions

The high resolution obtained in microspheres are due to two factors: a) The EM radiation transmitted through the object produces both evanescent waves which include information on the fine structures of the object (smaller than a wavelength), and propagating waves which include the large image of the object (with dimensions larger than a wavelength). The evanescent waves decay after distance of wavelength order, so that only the evanescent waves which are near the object are efficient in conserving the fine structure imaging. But one can attach the microsphere to a frame which moves on the object and scans a large image. b) By using boundary conditions according to complex Snell's law, we find that the component of a part of the evanescent waves perpendicular to the microsphere surface become real and propagating. While this effect is important there is another important factor which should be considered. The convolution between the EM waves, on the inner surface of the microsphere, and the transfer function enables propagation with small effective wavelengths (large wave vectors). This effect is due to spread of the wavevectors due to the above convolution. The transfer function includes also the wavevectors of the evanescent waves and thus enable them to be transferred to the image without evanescent wave decay. It is suggested to use Mie theory for calculating the transfer function, but such calculations are very complicated and usually give only numerical results. We suggest, therefore, to use the transfer function as experimental function which can be used also for the nano-jet production.

Figure 1 gives only a geometric optics picture. The incident and transmitted angles $\theta_I$ and $\theta_T$ angles are derived by Snell's law, respectively. There are different features in the microsphere system which can be explained by the geometric optics picture. For example, for small microsphere index of refraction we get diverging beam with virtual image, while if this index of refraction is large then the beam is converging with real image, like that in Figure 1. The angle $\beta$ between the beam converging to the nano-jet and the symmetric axis, and the distance $r$ from the point P to the symmetric axis can be obtained by simple geometric calculations. But the microsphere super-resolutions can be calculated, by using the analysis presented in the present article.

**Disclosures and declarations**

**Conflict of interest:** The author declares that there is no conflict of interest.

**Author contribution:** Investigation of Y. Ben-Aryeh

**Funding:** The present study was supported by Technion-Mossad under grant No. 2007256